\documentstyle[preprint,aps]{revtex}
\tightenlines
\input epsf.tex
\def\DESepsf(#1 width #2){\epsfxsize=#2 \epsfbox{#1}}
\begin{document}

\preprint{\vbox{\hbox{COLO-HEP-419}\hbox{hep-ph/9812530}}}
\draft
\title {Flavor SU(3) Symmetry and Factorization \\
in B Decays to Two Charmless Vector Mesons}
\author{Sechul Oh}
\address{Department of Physics, University of Colorado, Boulder, CO 80303}
\date{December, 1998}
\maketitle
\begin{abstract}
We first present a model-independent analysis using flavor SU(3) symmetry with SU(3) breaking effects in $B$ decays to two charmless vector mesons ($VV$) in the final state.  In order to bridge the flavor SU(3) symmetry approach and the factorization approach in $B \rightarrow VV$ decays, we explicitly show how to translate each SU(3) amplitude into a corresponding amplitude in factorization. 
Various decay modes, including $B \rightarrow K^* \phi$ which is so far the only $B \rightarrow VV$ mode having some experimental evidence, are discussed in both the SU(3) symmetry and factorization approaches.   
Within the generalized factorization approximation, the flavor SU(3) amplitudes and SU(3) breaking effects are numerically estimated as a function of the parameter $\xi \equiv 1/ N_c$ ($N_c$ the effective number of color).    
\end {abstract}

\newpage
The CLEO Collaboration\cite{1,2} has reported information on branching ratios of a number of exclusive decay modes where $B$ decays into a pair of pseudoscalars ($P$), a vector ($V$) and a pseudoscalar meson, or a pair of vector mesons.  
Some decay modes like $B^+ \rightarrow \omega \pi^+$, $B^+ \rightarrow \omega K^+$ and $B \rightarrow K^* \phi$ have been observed for the first time and in other decay modes improved bounds have been put.   
Motivated by this new information many works have been done in the framework of factorization approximation \cite{3,3a,300,301,302,303,4,5,500,6,7} to understand two-body $B$ decays such as $B \rightarrow PP$, $VP$ and $VV$, or in the context of flavor SU(3) symmetry (SU(3)$_F$) \cite{8,9,9a,10} to analyze $B \rightarrow PP$ and $VP$ decays.   

The generalized factorization approximation has been quite successfully used in
two body $D$ decays as well as $B\rightarrow D$ decays\cite{11}. The method 
includes color octet non-factorizable contribution by treating $\xi\equiv 1/N_c$ ($N_c$ denotes the effective number of color) as an adjustable parameter.  But, the validation of this approach in $B$ decays to light mesons such as 
$B \rightarrow PP, VP$ and $VV$ is not justified by experiment yet.   Furthermore, the numerical results like branching ratios obtained in this method is quite sensitive to several parameters involved in those decay, among which some parameters like a Cabibbo-Kobayashi-Maskawa (CKM) matrix element $V_{td}$ and certain form factors are not so far well known.  

On the other hand, there exists a different approach which is purely based on flavor SU(3) symmetry. 
In this method, the decay amplitudes of two body $B$ decays are decomposed into linear combinations of the SU(3)$_F$ amplitudes which are reduced matrix elements defined in Ref.\cite{8}.  
This approach, though more general, lacks detailed predictions that one can make using the generalized factorization.  

In this paper we attempt to bridge these different approaches in $B \rightarrow VV$ decays.  For this goal, we first present the SU(3)$_F$ analysis with broken SU(3)$_F$ symmetry in $B \rightarrow VV$ decays.   
Then each SU(3)$_F$ amplitude involved in the processes $B \rightarrow VV$ is \emph{translated into} a corresponding amplitude in factorization and SU(3)$_F$ breaking effects as well as each SU(3)$_F$ amplitude are numerically estimated as a function of the parameter $\xi$.  
Various decay modes are discussed in both the SU(3)$_F$ and factorization approaches.    

Since there are three possible partial waves in the final state, $B \rightarrow VV$ decays are more complicated than $B \rightarrow PP$ or $VP$ decays.  Three independent helicity amplitudes for the decay $B(p) \rightarrow V_1(p_1, \epsilon_1) V_2(p_2, \epsilon_2)$ are given by \cite{5,500}
\begin{eqnarray}
A_{\lambda} = < V_1(\lambda) V_2(\lambda) | H_{eff} | B >, \;\; (\lambda = 0, \pm 1)
\end{eqnarray}
where $\lambda$ is the helicity of the final-state vector mesons $V_1$ and $V_2$ in the rest frame of $B$ meson.  The helicity amplitudes can be expressed by three invariant amplitudes $a$, $b$, $c$, as follows: 
\begin{eqnarray}
A_{\lambda} = \epsilon^{\mu *}_1(\lambda) \epsilon^{\nu *}_2(\lambda) \left( a g_{\mu \nu} + {b \over m_1 m_2} p_{\mu} p_{\nu} + {ic \over m_1 m_2} \epsilon_{\mu \nu \alpha \beta} p^{\alpha}_1 p^{\beta} \right), 
\end{eqnarray}
where $p_i$ and $\epsilon_i$ are the four-momentum and polarization vector of the $V_i$ meson, respectively, and $p = p_1 +p_2$ is the four-momentum of the $B$ meson.  
For the SU(3)$_F$ analysis of $B \rightarrow VV$ decays, these partial waves in the final state need to be separated out.  We will assume that this can be done by certain methods such as one using angular distributions of $B$ decays\cite{12}.  

In SU(3)$_F$ decomposition of decay amplitudes of the $B \rightarrow VV$ modes, we choose the notations given in Refs.\cite{8,9,9a,10} as follows:     
we represent the decay amplitudes in terms of the basis of quark diagram contributions $T$ (tree), $C$ (color-suppressed tree), $P$ (QCD-penguin), $S$ (additional penguin effect involving SU(3)$_F$-singlet mesons), $E$ (exchange),  $A$ (annihilation), and $PA$ (penguin annihilation). 
(Here we have omitted the helicity index in all the above SU(3)$_F$ amplitudes; i.e., $T \equiv T_{\lambda}$, $C \equiv C_{\lambda}$ and so forth.  From now on, it is understood that the helicity index on the amplitudes is omitted unless it is specified.)
The amplitudes $E$, $A$ and $PA$ may be neglected to a good approximation because of a suppression factor of $f_B / m_B \approx 5 \%$. 
(It has been pointed out that in some decay modes these contributions may not be neglected\cite{7}.  We will remark this later.) 
For later convenience we also denote the electroweak (EW) penguin effects explicitly as $P_{EW}$ (color-favored EW penguin) and $P_{EW}^{C}$ (color-suppressed EW penguin), even though in terms of quark diagrams the inclusion of these EW penguin effects only leads to the following replacement without introducing new SU(3)$_F$ amplitudes : $T \rightarrow T + P_{EW}^{C}$, $C \rightarrow C + P_{EW}$, $P \rightarrow P -{1 \over 3} P_{EW}^{C}$, $S \rightarrow S -{1 \over 3} P_{EW}$.  
For SU(3)$_F$ breaking diagrams we use the notations defined in \cite{9}.     
The phase convention used for the vector mesons is  
\begin{eqnarray}
\rho^+ &=& u \bar d, \;\; \rho^0 = {1 \over \sqrt{2}} (d \bar d -u \bar u), \;\; 
\rho^- = - \bar u d,  \nonumber \\  
K^{*+} &=& u \bar s, \;\; K^{*0} = d \bar s, \;\; \bar K^{*0} = \bar d s, \;\; 
K^{*-} = - \bar u s,  \nonumber \\ 
\omega &=& {1 \over \sqrt{2}} (u \bar u + d \bar d), \;\; \phi = s \bar s.   
\end{eqnarray} 

In the factorization approach, we first consider the next-to-leading order effective Hamiltonian.  
We then use the generalized factorization approximation to derive hadronic matrix elements by saturating the vacuum state in all possible ways.  
The effective weak Hamiltonian for hadronic $\Delta B=1$ decays can be written as 
\begin{eqnarray}
 H_{eff} &=& {4 G_{F} \over \sqrt{2}} \left[ V_{ub}V^{*}_{uq} (c_1 O^{u}_1
+c_2 O^{u}_2) 
   + V_{cb}V^{*}_{cq} (c_1 O^{c}_1 +c_2 O^{c}_2)
   - V_{tb}V^{*}_{tq} \sum_{i=3}^{12} c_{i} O_{i} \right] \nonumber \\ 
  &+& h.c. ,
\end{eqnarray}  where $O_{i}$'s are defined as 
\begin{eqnarray}
O^f_1 &=& \bar q \gamma_{\mu} L f \cdot \bar f \gamma^{\mu} L b , \;\;
O^f_2 = \bar q_{\alpha} \gamma_{\mu} L f_{\beta} \cdot \bar f_{\beta}
\gamma^{\mu} L b_{\alpha},    \nonumber \\   
O_{3(5)} &=& \bar q \gamma_{\mu} L b \Sigma \bar q^{\prime} \gamma^{\mu} L(R)
q^{\prime},  \;\; 
O_{4(6)} = \bar q_{\alpha} \gamma_{\mu} L b_{\beta} \cdot \Sigma \bar
q^{\prime}_{\beta} \gamma^{\mu} L(R) q^{\prime}_{\alpha},  \nonumber \\ 
O_{7(9)} &=& {3 \over 2} \bar q \gamma_{\mu} L b \cdot \Sigma e_{q^{\prime}} \bar
q^{\prime} \gamma^{\mu} R(L) q^{\prime},  \;\;  
O_{8(10)} ={3 \over 2} \bar q_{\alpha} \gamma_{\mu} L b_{\beta} \cdot \Sigma
e_{q^{\prime}} \bar q^{\prime}_{\beta} \gamma^{\mu} R(L) q^{\prime}_{\alpha},  \nonumber \\
O_{11} &=& {g_s \over 32 \pi^2} m_b \bar q \sigma^{\mu \nu} R T^a b G^a_{\mu \nu},  \;\; 
O_{12} = {e \over 32 \pi^2} m_b \bar q_{\alpha} \sigma^{\mu \nu} R b_{\beta} F_{\mu \nu}. 
\end{eqnarray}  
Here $c_i$'s are the Wilson coefficients (WC's) evaluated at the renormalization scale $\mu$.  $L(R) = (1 \mp \gamma_5)/2$, $f$ can be $u$ or $c$ quark,
$q$ can be $d$ or $s$ quark, and $q^{\prime}$ is summed over $u$, $d$, $s$, and
$c$ quarks.  $\alpha$ and $\beta$ are the SU(3) color indices, and $T^a$ ($a= 1,...,8$) are the SU(3) generator with the normalization $Tr (T^a T^b) =\delta^{ab} /2$.  $g_s$ and $e$ are the strong and electric coupings, respectively.  $G^a_{\mu \nu}$ and $F_{\mu \nu}$ denote the gluonic and photonic field strength tensors, respectively.  
$O_1$, $O_2$ are the tree-level and QCD-corrected operators.  
$O_{3-6}$ are the gluon-induced strong penguin operators.  $O_{7-10}$ are the
EW penguin  operators due to $\gamma$ and $Z$ exchange, and box diagrams at loop level.  We shall take into account the chromomagnetic operator $O_{11}$, but neglect the small contribution from $O_{12}$ \cite{6,7}. 

First, we list the $B \rightarrow VV$ decay modes in terms of the SU(3)$_F$ amplitudes.  The coefficients of SU(3)$_F$ amplitudes in $B \rightarrow VV$ are listed in Tables I and III for strangeness-conserving ($\Delta S =0$) and strangeness-changing ($|\Delta S| =1$) processes, respectively.  The SU(3)$_F$ breaking effects are shown in Tables II and IV. 
Here we use the notations for SU(3)$_F$ breaking diagrams relevant to the singlet contribution $S^{(\prime)}$ in the similar way to the $P^{(\prime)}$ case.  In $B \rightarrow K^* \phi$ the SU(3)$_F$ breaking effects are expected to be large and are denoted as $P^{\prime V}_{(3)}$ and so on.    
In the tables, unprimed and primed letters denote $\Delta S =0$ and $|\Delta S| =1$ processes, respectively.  For later purpose the superscript, $V_1$ or $V_2$, on each SU(3)$_F$ amplitude is used to describe which vector meson ($V_1$ or $V_2$) includes the spectator quark in the corresponding quark diagram.  
In SU(3)$_F$ limit the SU(3)$_F$ amplitudes with the same quark diagram contribution but the superscript are in fact equivalent: i.e., $T^{(\prime)V_1} = T^{(\prime)V_2} \equiv T^{(\prime)V}$ and so forth.  
Thus, for instance, in SU(3)$_F$ limit the amplitude of the decay $B^0 \rightarrow \rho^0 \omega$ has only penguin contributions since $(1/2) C^{V_1}+(-1/2) C^{V_2} =0$. 

Among the $\Delta S =0$ amplitudes the tree contribution $T^V$ is expected to be largest so that the decays $B^+ \rightarrow \rho^+ \rho^0$, $\rho^+ \omega$ and 
$B^0 \rightarrow \rho^+ \rho^-$ are expected to have the largest rates. 
The decay modes $B^+ \rightarrow \rho^+ \phi$, $K^{*+} \bar K^{*0}$ and $B^0 \rightarrow \rho^0 \phi$, $\omega \phi$, $K^{*0} \bar K^{*0}$, $\rho^0 \omega$ have only penguin contributions and are expected to have smaller decay rates. 
In particular, since $B^+ \rightarrow \rho^+ \phi$ and $B^0 \rightarrow \rho^0 \phi$, $\omega \phi$ decays have only the singlet and EW penguin contributions ($S^V - {1 \over 3} P^V_{EW}$), the decay rates of these modes will be very small because of the Okubo-Zweig-Iizuka (OZI) suppression (see below).   
However this is not the case for the $|\Delta S| =1$ decays with only penguin contributions such as $B^+ \rightarrow K^{*0} \rho^+$, $K^{*+} \phi$ and $B^0 \rightarrow K^{*0} \phi$, since the branching ratios for those decay modes can be enhanced by the ratio of the CKM elements $|V^*_{tb} V_{ts}| / |V^*_{ub} V_{us}| \approx 50$.
In the $|\Delta S|=1$ decays, the amplitude $P^{\prime V}$ is expected to be the dominant contribution.  The singlet amplitude $S^{\prime V}$ is expected to be smaller and may be relatively unimportant\cite{10}; like the case of $B \rightarrow VP$ (e.g., $B \rightarrow \omega K$ and $\phi K$) this amplitude would involve a coupling of the final-state vector mesons $\omega$ and $\phi$, and it violates the OZI rule which favors connected quark diagrams.   In $B \rightarrow VP$ decays the similar amplitude $s^{\prime}_V (\equiv S^{\prime}_V -{1 \over 3} P^{\prime}_{EW,V})$ defined in Ref.\cite{10} contributes only to decays involving the pseudoscalar mesons $\eta$ and $\eta^{\prime}$ and is expected not to be very small since the flavor-singlet couplings of the $\eta$ and $\eta^{\prime}$ can be affected by the axial anomaly\cite{1200}.  

However, in the framework of factorization, the contribution of $S^{\prime V}$ strongly depends on $\xi$ and could be comparable to the contribution of $P^{\prime V}$ for some values of $\xi$ as shown in Table V.  This $\xi$ dependence arises from the fact that $S^{\prime V}$ is proportional to the effective coefficient ($a_3 +a_5$) as in Eq.(\ref{su3facto}) below.  The value of the effective coefficient ($a_3 +a_5$) changes very sensitively depending on $\xi$; for example, from $-(187+28 i) \times 10^{-4}$ at $\xi =0.5$, to $+3 \times 10^{-4}$ at $\xi =1/3$, and to $+(383+58 i) \times 10^{-4}$ at $\xi =0$ \cite{7}.  We note that the contribution of $S^{\prime V}$ is highly suppressed at near $\xi=1/3$.  The amplitude $P^{\prime V}$ is proportional to the effective constant $a_4$ as in Eq.(\ref{su3facto}) and the value of $a_4$ is much less sensitive to $\xi$; for instance,  from $-(402+72 i) \times 10^{-4}$ at $\xi =0.5$, to $-(439+77 i) \times 10^{-4}$ at $\xi =1/3$, and to $-(511+87 i) \times 10^{-4}$ at $\xi =0$.  Thus one can expect in the factorization approach that in some decay modes which have both of the penguin contributions $P^{\prime V}$ and $S^{\prime V}$, these two contributions interfere constructively or destructively, depending on $\xi$;  the interference is constructive for $\xi=0.5$, but the interference is destructive for $\xi=1/3$ and $\xi=0$.  In particular, for $\xi=0$, the destructive interference is so large that the relevant decay rate is expected to be reduced in a large amount.  This is the case of the process $B \rightarrow K^* \phi$\cite{annih}, which is the only $B \rightarrow VV$ mode having some experimental evidence so far.  The CLEO Collaboration has reported an averaged branching ratio $BR (B \rightarrow K^* \phi) =(1.1^{+0.6}_{-0.5}\pm 0.2) \times 10^{-5}$ when charged and neutral modes are combined\cite{2}.  The amplitudes of decays $B^+ \rightarrow K^{*+} \phi$ and $B^0 \rightarrow K^{*0} \phi$ can be expressed as $P^{\prime V} +S^{\prime V} -{1 \over 3} P^{\prime C, V}_{EW} -{1 \over 3} P^{\prime V}_{EW}$ (see Table III).  So in factorization the branching ratios of these modes are expected to have the smallest values at $\xi=0$ (see Figure 1), which are in fact incompatible with the experimental data.  It has been shown that the experimentally favored values of $\xi$ are $0.4 \leq \xi \leq 0.6$, which are quite different from the values of $\xi$ implied by analyses of $B \rightarrow VP$ decays\cite{4,7}.  
Improved measurements of the branching ratios of the decay modes $B^+ \rightarrow K^{*+} \phi$ and $B^0 \rightarrow K^{*0} \phi$ will provide crucial information for test of the factorization ansatz.   

One can test the OZI suppression of the $S^{\prime V}$ amplitude by measuring the branching ratios of $B^+ \rightarrow K^{*+} \rho^0$ and $B^+ \rightarrow K^{*+} \omega$ decays.  The amplitudes of these processes can be described as $- {1 \over \sqrt{2}} (T^{\prime V} + C^{\prime V} +P^{\prime V} +P^{\prime V}_{EW} +{2 \over 3} P^{\prime C, V}_{EW})$ and ${1 \over \sqrt{2}} (T^{\prime V} + C^{\prime V} +P^{\prime V} +2 S^{\prime V} +{1 \over 3} P^{\prime V}_{EW} +{2 \over 3} P^{\prime C, V}_{EW})$, respectively.  
So (in nonet symmetry) the branching ratios of both decays are expected to be similar, if $S^{\prime V}$ is highly suppressed.  
In addition, if the contribution of $T^{\prime V}$ and $C^{\prime V}$ is much smaller than that of $P^{\prime V}$, the branching ratios of $B^+ \rightarrow K^{*+} \rho^0$ and $B^+ \rightarrow K^{*+} \omega$ will be of the same order as that of $B^+ \rightarrow K^{*0} \rho^+$ whose amplitude is only $P^{\prime V}-{1 \over 3} P^{\prime C, V}_{EW}$.  
In the $B^+ \rightarrow K^{*+} \phi$ mode, the SU(3)$_F$ breaking effect is expected to be sizable; naively, $(f_{\phi} m_{\phi})/ (f_{\rho} m_{\rho}) \approx 1.24$.  After obtaining the information about the $S^{\prime V}$ suppression, one may measure the SU(3)$_F$ breaking by comparing the rates of decays $B^+ \rightarrow K^{*+} \phi$ and $B^+ \rightarrow K^{*0} \rho^+$.     

Each SU(3)$_F$ amplitude can be measured through certain decay processes.                         In $\Delta S=0$ decays, the strong penguin contribution $P^V$ (combined with \emph{very small} color-suppressed EW penguin contribution) is measured in $B^{+(0)} \rightarrow K^{*+(0)} \bar K^{*0}$\cite{annih}.  The singlet contribution (combined with color-favored EW penguin) $s^V \equiv S^V -(1/3) P_{EW}^{V}$ is measured in $B^{+(0)} \rightarrow \rho^{+(0)} \phi$ or $B^0 \rightarrow \omega \phi$, even though the branching ratios of these modes seem to be much smaller than those of the dominant modes $B^+ \rightarrow \rho^+ \rho^0$ and $\rho^+ \omega$.    
The dominant contribution $T^V$ is measured in $B^0 \rightarrow \rho^+ \rho^-$ and the color-suppressed contribution $C^V$ is measured by the combination $\sqrt{2} A(B^+ \rightarrow \rho^+ \rho^0) -A(B^0 \rightarrow \rho^+ \rho^-)$.  (Here the helicity index $\lambda$ is omitted.)   
Similarly, in $|\Delta S|=1$ decays, the dominant contribution $P^{\prime V}$ is measured in $B^+ \rightarrow K^{*0} \rho^+$ and the singlet contribution (combined with color-favored EW penguin) $s^{\prime V} \equiv S^{\prime V} -(1/3) P_{EW}^{\prime V}$ is measured by the combination $A(B^+ \rightarrow K^{*+} \rho^0) + A(B^+ \rightarrow K^{*0} \omega)$.  
The tree (combined with color-suppressed EW penguin) $t^{\prime V} \equiv T^{\prime V}+P^{\prime C, V}_{EW}$ and color-suppressed (combined with color-favored EW penguin) $c^{\prime V} \equiv C^{\prime V}+P^{\prime V}_{EW}$ are measured by the combinations $A(B^0 \rightarrow K^{*+} \rho^-) +A(B^+ \rightarrow K^{*0} \rho^+)$ and $A(B^+ \rightarrow K^{*0} \rho^+) - \sqrt{2} A(B^0 \rightarrow K^{*0} \rho^0)$, respectively.    

From Tables I$-$IV, one can easily find some useful relations among the decay amplitudes.  (The helicity index on the amplitudes is omitted.)  
The equivalence relations are    
\begin{eqnarray}
A( B^+ \rightarrow \rho^+ \phi )&=& -\sqrt{2} A( B^0 \rightarrow \rho^0 \phi ), \\ 
A( B^+ \rightarrow K^{*+} \phi )&=& A( B^0 \rightarrow K^{*0} \phi ), \\
A( B^+ \rightarrow K^{*+} \bar K^{*0} )&=& A(B^0 \rightarrow K^{*0} \bar K^{*0}).  \end{eqnarray} 
The triangle and quadrangle relations are : \\
for $\Delta S=0$ decays, 
\begin{eqnarray}
A( B^0 \rightarrow \rho^+ \rho^- ) &+& \sqrt{2} A( B^0 \rightarrow \rho^0 \rho^0 ) = \sqrt{2} A( B^+ \rightarrow \rho^+ \rho^0 ), 
\label{tri1} \\
A( B^0 \rightarrow \rho^+ \rho^- ) &+& \sqrt{2} A( B^+ \rightarrow \rho^+ \omega ) = \sqrt{2} A( B^0 \rightarrow \omega \omega ), 
\label{tri2} \\ 
\sqrt{2} A( B^+ \rightarrow \rho^+ \rho^0 ) &+& \sqrt{2} A( B^+ \rightarrow \rho^+ \omega ) = 2 A( B^0 \rightarrow \rho^0 \omega )   \nonumber \\
&=& \sqrt{2} A( B^0 \rightarrow \rho^0 \rho^0 ) + \sqrt{2} A( B^0 \rightarrow \omega \omega )  \nonumber \\
&=& 2 A( B^+ \rightarrow \rho^+ \phi ) + 2 A( B^+ \rightarrow K^{*+} \bar K^{*0}),  
\label{tri3}
\end{eqnarray}  
and for $|\Delta S|=1$ decays, 
\begin{eqnarray} 
\sqrt{2} A( B^0 \rightarrow K^{*0} \rho^0) &+& \sqrt{2} A( B^0 \rightarrow K^{*0} \omega) = 2 A( B^+ \rightarrow K^{*+} \phi) = 2 A( B^0 \rightarrow K^{*0} \phi),  \\
\sqrt{2} A( B^+ \rightarrow K^{*+} \rho^0) &+& A( B^+ \rightarrow K^{*0} \rho^+) = A( B^0 \rightarrow K^{*+} \rho^-) +\sqrt{2} A( B^0 \rightarrow K^{*0} \rho^0), \label{tri4} \\
\sqrt{2} A( B^+ \rightarrow K^{*+} \rho^0) &+& 2 A( B^+ \rightarrow K^{*0} \rho^+) = 2 A( B^+ \rightarrow K^{*+} \phi) - \sqrt{2} A( B^+ \rightarrow K^{*+} \omega),  \\
A( B^+ \rightarrow K^{*0} \rho^+) &+& A( B^0 \rightarrow K^{*+} \rho^-) 
= \sqrt{2} A( B^0 \rightarrow K^{*0} \omega) - \sqrt{2} A( B^+ \rightarrow K^{*+} \omega). 
\end{eqnarray}

As in the cases of $B \rightarrow PP$ and $VP$, these SU(3)$_F$ relations can be used in various ways.  One obvious application of these relations is to serve as tests of flavor SU(3) symmetry with comparisons of magnitudes of decay amplitudes involved in each \emph{polygon} relations.  These relations may also be used to predict the amplitudes and phases of some of the decay processes from others by constructing the relevant polygons in the same plane.  For example, suppose the amplitudes of the decays $B^+ \rightarrow \rho^+ \rho^0$, $\rho^+ \omega$ and $B^0 \rightarrow \rho^+ \rho^-$, $\rho^0 \rho^0$, $\omega \omega$ are known from experimental measurement.  
Using the relations (\ref{tri1}) and (\ref{tri2}), two triangles with the same base $|A( B^0 \rightarrow \rho^+ \rho^- )|$ can be constructed in the same plane.  Then by constructing the third triangle of the relation (\ref{tri3}), one can determine $|A( B^0 \rightarrow \rho^0 \omega )|$ and the relative phases among the amplitudes of the decays $B^+ \rightarrow \rho^+ \rho^-$, $\rho^+ \omega$ and $B^0 \rightarrow \rho^0 \omega$ up to two-fold ambiguity.  
The ratios of magnitudes of CKM matrix elements can be determined by comparing decay rates between certain $\Delta S=0$ and $|\Delta S|=1$ modes, such as 
\begin{eqnarray}
{ BR(B^{+(0)} \rightarrow K^{*+(0)} \bar K^{*0}) \over BR(B^+ \rightarrow K^{*0} \rho^+) } = \left| { V_{td} \over V_{ts} } \right|^2.   
\end{eqnarray}
Here the top quark dominance in the penguin contributions is assumed.  

Now we \emph{translate} each SU(3)$_F$ amplitude involved in $B \rightarrow VV$ decays \emph{into} the explicit form calculated in the factorization approximation.   
We find for $B \rightarrow V_1 V_2$
\begin{eqnarray}
T^{(\prime) V_j} &=& V^*_{ub} V_{ud (s)} Y^{V_j} (- a_1),  \ \ \ \ \ \ \
C^{(\prime) V_j} = V^*_{ub} V_{ud (s)} Y^{V_j} (- a_2),  \nonumber \\ 
S^{(\prime) V_j} &=& V^*_{tb} V_{td (s)} Y^{V_j} (a_3 +a_5),  \ \ \ \ 
P^{(\prime) V_j} = V^*_{tb} V_{td (s)} Y^{V_j} a_4,  \nonumber \\ 
P_{EW}^{(\prime) V_j} &=& V^*_{tb} V_{td (s)} Y^{V_j} {3 \over 2} (a_7 +a_9), 
\ \
P_{EW}^{(\prime) C, V_j} = V^*_{tb} V_{td (s)} Y^{V_j} {3 \over 2} a_{10}, \ \
\label{su3facto}
\end{eqnarray} 
where 
\begin{eqnarray}
Y^{V_j} &\equiv& i {G_F \over \sqrt{2}} f_{V_k} m_{V_k} F^{B \rightarrow V_j}, \ \ (j \neq k; \; j, k = 1,2)  \\
F^{B \rightarrow V_j} &\equiv&  i \epsilon^{\mu}_{V_k} < V_j(p_{V_j})| V_{\mu}-A_{\mu} | B(p_B) >  \nonumber \\ 
&=& - i \epsilon_{\mu \nu \alpha \beta} \epsilon^{\mu}_{V_j} \epsilon^{\nu}_{V_k} p^{\alpha}_B p^{\beta}_{V_j} {2 V^{B \rightarrow V_j} (m^2_{V_k}) \over (m_B +m_{V_j})}
+ (\epsilon_{V_j} \cdot \epsilon_{V_k}) 
(m_B +m_{V_j}) A_1^{B \rightarrow V_j} (m^2_{V_k})  \nonumber \\
&\mbox{}& - (\epsilon_{V_j} \cdot p_B) 
(\epsilon_{V_k} \cdot p_B) {2 A_2^{B \rightarrow V_j} (m^2_{V_k}) \over 
(m_B +m_{V_j})}.
\end{eqnarray}
Here the effective coefficients $a_i$ are defined as $a_i = c^{eff}_i + \xi c^{eff}_{i+1}$ ($i =$ odd) and $a_i = c^{eff}_i + \xi c^{eff}_{i-1}$ ($i =$ even) with the effective WC's $c^{eff}_i$ at the scale $m_b$.  $m_{B(V)}$ and $p_{B(V)}$ are the mass and four-momentum of $B$ meson (a vector meson $V$), and $\epsilon_V$ is the polarization vector of $V$.  $f_V$, $V^{B \rightarrow V}$ and $A^{B \rightarrow V}_{1,2}$ are the decay constant and form factors of $V$, respectively.  

With Tables I, III and the above relations (\ref{su3facto}), one can easily write down in the factorization approximation the amplitude of \emph{any} $B \rightarrow VV$ decay mode in the tables.  For example, from Table I and the relations (\ref{su3facto}), the amplitude of the process $B^+ \rightarrow \rho^+ \omega$ can be written as  
\begin{eqnarray}
A(B^+ \rightarrow \rho^+ \omega) 
&=& {1 \over \sqrt{2}} \left[ ( C^{\rho} + P^{\rho} +2 S^{\rho} +{1 \over 3} P^{\rho}_{EW} -{1 \over 3} P^{C,\rho}_{EW} ) + ( T^{\omega} +P^{\omega} +{2 \over 3} P^{C,\omega}_{EW} ) \right]    \nonumber \\
&=& -i {G_F \over 2} f_{\omega} m_{\omega} \left[ - i \epsilon_{\mu \nu \alpha \beta} \epsilon^{\mu}_{\omega} \epsilon^{\nu}_{\rho} p^{\alpha}_B p^{\beta}_{\rho} {2 V^{B \rightarrow \rho} (m^2_{\omega}) \over (m_B +m_{\rho})}   \right. \nonumber \\
&\mbox{}& \left.  + (\epsilon_{\rho} \cdot \epsilon_{\omega}) 
(m_B +m_{\rho}) A_1^{B \rightarrow \rho} (m^2_{\omega})
- (\epsilon_{\rho} \cdot p_B) 
(\epsilon_{\omega} \cdot p_B) {2 A_2^{B \rightarrow \rho} (m^2_{\omega}) \over 
(m_B +m_{\rho})}  \right]  \nonumber \\
&\mbox{}& \cdot \left[ V^*_{ub} V_{ud} a_2 - V^*_{tb} V_{td} (a_4 + 2(a_3 +a_5) +{1 \over 2} (a_7 +a_9 -a_{10}) \right]   \nonumber \\
&\mbox{}& -i {G_F \over 2} f_{\rho} m_{\rho} \left[ - i \epsilon_{\mu \nu \alpha \beta} \epsilon^{\mu}_{\rho} \epsilon^{\nu}_{\omega} p^{\alpha}_B p^{\beta}_{\omega} {2 V^{B \rightarrow \omega} (m^2_{\rho}) \over (m_B +m_{\omega})}  \right. \nonumber \\
&\mbox{}& \left.  + (\epsilon_{\omega} \cdot \epsilon_{\rho}) 
(m_B +m_{\omega}) A_1^{B \rightarrow \omega} (m^2_{\rho})
- (\epsilon_{\omega} \cdot p_B) 
(\epsilon_{\rho} \cdot p_B) {2 A_2^{B \rightarrow \omega} (m^2_{\rho}) \over 
(m_B +m_{\omega})}  \right]  \nonumber \\
&\mbox{}& \cdot \left[ V^*_{ub} V_{ud} a_1 - V^*_{tb} V_{td} (a_4 + a_{10}) \right]. 
\end{eqnarray}

Decay rates of $B \rightarrow VV$ have been calculated in the framework of factorization for several years\cite{3,5,500,7}.  Those calculations have used specific $B \rightarrow V$ form factors which are model-dependent.  
We make a numerical estimation on magnitudes of each SU(3)$_F$ amplitudes and SU(3)$_F$ breaking effects in the factorization approximation, using the expressions in Eq. (\ref{su3facto}).  
We use two sets of form factors: one based on Bauer, Stech, and Wirbel (BSW) quark model\cite{13}, and the other based on lattice QCD and light-cone QCD sum rules\cite{14}.  
The form factors are: (i) in the BSW model, $V^{B \rightarrow \rho (\omega)}(0)= 0.33$, $A_{1,2}^{B \rightarrow \rho (\omega)}(0)= 0.28$, $V^{B \rightarrow K^*}(0)= 0.37$, $A_{1,2}^{B \rightarrow K^*}(0)= 0.33$, (ii) based on lattice QCD and QCD sum rules, $V^{B \rightarrow \rho (\omega)}(0)= 0.35 \pm 0.05$, $A_1^{B \rightarrow \rho (\omega)}(0)= 0.27 \pm 0.04$, $A_2^{B \rightarrow \rho (\omega)}(0)= 0.26 \pm 0.04$, $V^{B \rightarrow K^*}(0)= 0.48 \pm 0.09$, $A_1^{B \rightarrow K^*}(0)= 0.35 \pm 0.07$, $A_2^{B \rightarrow K^*}(0)= 0.34 \pm 0.06$.  
The values of the decay constants (in MeV) we use are\cite{7,11}: $f_{\rho}=210$, $f_{\omega}=195$, $f_{K^*}=214$, $f_{\phi}=233$.  
We use numerical values of effective WC's at the scale $m_b$ given in Ref.\cite{7}.  The results are shown in Tables V and VI.  Here we have calculated magnitudes of the SU(3)$_F$ amplitudes and SU(3)$_F$ breaking effects averaged over helicity states for $\xi=0, \; 0.2, \; 1/3$, and 0.5.  For example, the magnitude of the tree amplitude $|T^{(\prime) V}|$ is given by 
\begin{eqnarray}
|T^{(\prime) V}| 
\equiv \left( {1 \over 3} \sum_{\lambda} |T^{(\prime) V_j}|^2 \right)^{1/2} 
&=&  {G_F \over \sqrt{2}} 
|V^*_{ub} V_{ud (s)}| (f_{V_k} m_{V_k}) 
 \left[ |a|^2 (2+x^2) +|b|^2 (x^2-1)^2  \right. \nonumber \\
&+& |c|^2 2 (x^2-1) +2 Re(ab^*) x (x^2-1) \left. \right]^{1/2} |a_1| , 
\label{Tv}
\end{eqnarray}
where 
\begin{eqnarray}
a &=& (m_B +m_{V_j}) A_1^{B \rightarrow V_j} (m^2_{V_k}),  \;\;\; 
b = - {2 m_{V_j} m_{V_k} \over m_B +m_{V_j}} 
 A_2^{B \rightarrow V_j}(m^2_{V_k}),  \nonumber \\
c &=& - {2 m_{V_j} m_{V_k} \over m_B +m_{V_j}} 
 V^{B \rightarrow V_j}(m^2_{V_k}),  \;\;\; 
x = { p_{V_j} p_{V_k} \over m_{V_j} m_{V_k}} 
 = { m^2_B -m^2_{V_j} -m^2_{V_k} \over 2 m_{V_j} m_{V_k} }.  
\label{abcx}
\end{eqnarray}

In Table V we see, as expected, that the tree contribution $T^{V}$ and the penguin contribution $P^{\prime V}$ dominate in $\Delta S=0$ and $|\Delta S|=1$ decays, respectively, independent of $\xi$.  
Among the SU(3)$_F$ amplitudes, the singlet contribution $S^{(\prime) V}$ and the color-suppressed EW penguin contribution $P^{(\prime) C,V}_{EW}$ are most sensitive to $\xi$.  Since the $P^{(\prime) C,V}_{EW}$ contribution is usually very small, in factorization the decay rates which include the the $P^{(\prime) C,V}_{EW}$ contribution are not actually affected by $\xi$ dependence of $P^{(\prime) C,V}_{EW}$.  However, as we have pointed out above, the $S^{\prime V}$ contribution changes very sensitively depending on $\xi$ (e.g., by about two orders in magnitude) and are responsible for strong $\xi$ dependence of certain decay rates like $B^{+(0)} \rightarrow K^{*+(0)} \phi$.  We see in Figure 1 that the combined contribution $P^{\prime V} + S^{\prime V}$ monotonically increases as $\xi$ increases.  
Similarly, since the $S^V$ contribution is very sensitive to $\xi$, decay rates of some $\Delta S=0$ processes such as $B^+ \rightarrow \rho^+ \phi$, $B^0 \rightarrow \rho^0 \phi$ and $\omega \phi$ are strongly dependent on $\xi$ (e.g., from $O(10^{-7})$ at $\xi=0$ to $O(10^{-9})$ at $\xi=1/3$).  
The SU(3)$_F$ breaking effects in the amplitudes arise from the decay constants, the form factors, and masses of vector mesons, as one can see in Eqs. (\ref{Tv}) and (\ref{abcx}).  For example, taking into account the SU(3)$_F$ breaking effects, the ratio of $T^{\prime V}$ and $T^V$ is estimated by $|T^{\prime V} /T^V|                                                                                                                                                                                                                                                                                                                       \approx \lambda \cdot (f_{K*} m_{K*}) /(f_{\rho} m_{\rho})$, where $\lambda= V_{us}$.   
An estimation of the SU(3)$_F$ breaking effects is shown in Table VI.  The breaking effects $P^{\prime V}_{(3)}$ and $S^{\prime V}_{(3)}$ are relatively large, since in the final state of $B \rightarrow K^* \phi$ there are \emph{three} (anti) $s$-quarks whose mass difference from the $d$-quark mass breaks SU(3)$_F$. 

In conclusion, we have presented a model-independent analysis based on broken flavor SU(3) symmetry in $B \rightarrow VV$ decays.  Each SU(3)$_F$ amplitude involved in $B \rightarrow VV$ has been translated into the explicit form in factorization and the SU(3)$_F$ breaking effects as well as each SU(3)$_F$ amplitude have been numerically estimated as a function of the parameter $\xi$.  
Various decay modes have been discussed in both the SU(3)$_F$ and factorization approaches.
Purely based on SU(3)$_F$ analysis, we have shown that in $\Delta S=0$ decays certain decay modes such as $B^+ \rightarrow \rho^+ \rho^0$ and $\rho^+ \omega$  are expected to have the largest decay rates and so they can be preferable modes to find in future experiment.  In $B \rightarrow K^* \phi$ which is so far the only $B \rightarrow VV$ mode having some experimental evidence and an averaged branching ratio $BR (B \rightarrow K^* \phi) =(1.1^{+0.6}_{-0.5}\pm 0.2) \times 10^{-5}$\cite{2}, we have discussed the singlet contribution $S^{\prime V}$ in detail in both the SU(3)$_F$ approach and the factorization approach, and emphasized the strong $\xi$ dependence of $S^{\prime V}$ which hinders a stable prediction of branching ratio for $B \rightarrow K^* \phi$ in factorization.  We have also suggested a test of suppression of the $S^{\prime V}$ contribution in experiment by measuring the branching ratios of $B^+ \rightarrow K^{*+} \rho^0$ and $B^+ \rightarrow K^{*+} \omega$ decays.  
To test the factorization ansatz, improved measurements of the branching ratios of the decays $B^+ \rightarrow K^{*+} \phi$ and $B^0 \rightarrow K^{*0} \phi$ as well as measurements of other decay modes will be rather crucial.   \\   

We would like to thank K.T. Mahanthappa and B. Dutta for helpful discussions.  Also, we would like to be grateful to N.G. Deshpande, Jim Smith, and Xiao-Gang He for critically reading the manuscript and giving helpful comments.  
This work was supported in part by the US Department of Energy Grant No. DE FG03-95ER40894.

\newpage 
\begin{table}
\caption{Coefficients of SU(3)$_F$ amplitudes in $B \rightarrow V_1 V_2$ ( $\Delta S = 0$ ).}
\begin{tabular}{|c||c|c|c|c|c|c|c|c|c|c|c|c|} 
$B \rightarrow V_1 V_2$ & $T^{V_1}$ & $T^{V_2}$ & $C^{V_1}$ & $C^{V_2}$ & $P^{V_1}$ & $P^{V_2}$ & $S^{V_1}$ & $S^{V_2}$ & $P_{EW}^{V_1}$ & $P_{EW}^{V_2}$ & $P_{EW}^{C,V_1}$ & $P_{EW}^{C,V_2}$ \\ \hline \hline
$B^+ \rightarrow \rho^+ \rho^0$ & 0 & $-{1 \over \sqrt{2}}$ & $-{1 \over \sqrt{2}}$ & 0 & ${1 \over \sqrt{2}}$ & $-{1 \over \sqrt{2}}$ & 0 & 0 & $-{1 \over \sqrt{2}}$ & 0 & $-{1 \over 3\sqrt{2}}$ & $-{2 \over 3\sqrt{2}}$  \\ 
$B^+ \rightarrow \rho^+ \omega$ & 0 & $1 \over \sqrt{2}$ & $1 \over \sqrt{2}$ & 0 & $1 \over \sqrt{2}$ & $1 \over \sqrt{2}$ & $\sqrt{2}$ & 0 & $1 \over 3\sqrt{2}$ & 0 & $-{1 \over 3\sqrt{2}}$ & $2 \over 3\sqrt{2}$  \\ 
$B^+ \rightarrow \rho^+ \phi$ & 0 & 0 & 0 & 0 & 0 & 0 & 1 & 0 & $-{1 \over 3}$ & 0 & 0 & 0 \\ 
$B^+ \rightarrow K^{*+} \bar K^{*0} $ & 0 & 0 & 0 & 0 & 1 & 0 & 0 & 0 & 0 & 0 & $-{1 \over 3}$ & 0 \\ \hline 
$B^0 \rightarrow \rho^+ \rho^-$ & 0 & $-1$ & 0 & 0 & 0 & $-1$ & 0 & 0 & 0 & 0 & 0 & $-{2 \over 3}$ \\ 
$B^0 \rightarrow \rho^0 \rho^0$ & 0 & 0 & $-{1 \over 2\sqrt{2}}$ & $-{1 \over 2\sqrt{2}}$ & ${1 \over 2\sqrt{2}}$ & ${1 \over 2\sqrt{2}}$ & 0 & 0 & $-{1 \over 2\sqrt{2}}$ & $-{1 \over 2\sqrt{2}}$ & $-{1 \over 6\sqrt{2}}$ & $-{1 \over 6\sqrt{2}}$ \\ 
$B^0 \rightarrow \rho^0 \omega$ & 0 & 0 & $1 \over 2$ & $-{1 \over 2}$ & $1 \over 2$ & $1 \over 2$ & 1 & 0 & $1 \over 6$ & $-{1 \over 2}$ & $-{1 \over 6}$ & $-{1 \over 6}$ \\ 
$B^0 \rightarrow \rho^0 \phi$ & 0 & 0 & 0 & 0 & 0 & 0 & $1 \over \sqrt{2}$ & 0 & $-{1 \over 3\sqrt{2}}$ & 0 & 0 & 0 \\ 
$B^0 \rightarrow \omega \omega$ & 0 & 0 & ${1 \over 2\sqrt{2}}$ & ${1 \over 2\sqrt{2}}$ & ${1 \over 2\sqrt{2}}$ & ${1 \over 2\sqrt{2}}$ & ${1 \over \sqrt{2}}$ & ${1 \over \sqrt{2}}$ & ${1 \over 6\sqrt{2}}$ & ${1 \over 6\sqrt{2}}$ & $-{1 \over 6\sqrt{2}}$ & $-{1 \over 6\sqrt{2}}$ \\ 
$B^0 \rightarrow \omega \phi$ & 0 & 0 & 0 & 0 & 0 & 0 & $1 \over \sqrt{2}$ & 0 & $-{1 \over 3\sqrt{2}}$ & 0 & 0 & 0 \\ 
$B^0 \rightarrow K^{*0} \bar K^{*0} $ & 0 & 0 & 0 & 0 & 1 & 0 & 0 & 0 & 0 & 0 & $-{1 \over 3}$ & 0 
\end{tabular}
\vspace{2cm}
\caption{Coefficients of SU(3)$_F$ breaking effects in $B \rightarrow V_1 V_2$ ( $\Delta S = 0$ ).}
\begin{tabular}{|c||c|c|c|c|} 
 $B \rightarrow V_1 V_2$ & $P^{V_1}_3$ & $S^{V_1}_3$ & $P_{EW,3}^{V_1}$ & $P_{EW,3}^{C,V_1}$  \\ \hline \hline
$B^+ \rightarrow \rho^+ \phi$ & 0 & 1 & $-{1 \over 3}$ & 0  \\ 
$B^+ \rightarrow K^{*+} \bar K^{*0}$ & 1 & 0 & 0 & $-{1 \over 3}$  \\ \hline 
$B^0 \rightarrow \rho^0 \phi$ & 0 & $1 \over \sqrt{2}$ & $-{1 \over 3\sqrt{2}}$ & 0  \\ 
$B^0 \rightarrow \omega \phi$ & 0 & $1 \over \sqrt{2}$ & $-{1 \over 3\sqrt{2}}$ & 0  \\ 
$B^0 \rightarrow K^{*0} \bar K^{*0} $ & 1 & 0 & 0 & $-{1 \over 3}$ 
\end{tabular}
\newpage 
\caption{Coefficients of SU(3)$_F$ amplitudes in $B \rightarrow V_1 V_2$ ( $|\Delta S| = 1$ ).}
\begin{tabular}{|c||c|c|c|c|c|c|c|c|c|c|c|c|} 
 $B \rightarrow V_1 V_2$ & $T^{\prime V_1}$ & $T^{\prime V_2}$ & $C^{\prime V_1}$ & $C^{\prime V_2}$ & $P^{\prime V_1}$ & $P^{\prime V_2}$ & $S^{\prime V_1}$ & $S^{\prime V_2}$ & $P_{EW}^{\prime V_1}$ & $P_{EW}^{\prime V_2}$ & $P_{EW}^{\prime C,V_1}$ & $P_{EW}^{\prime C,V_2}$ \\ \hline \hline
$B^+ \rightarrow K^{*+} \rho^0$ & 0 & $-{1 \over \sqrt{2}}$ & $-{1 \over \sqrt{2}}$ & 0 & 0 & $-{1 \over \sqrt{2}}$ & 0 & 0 & $-{1 \over \sqrt{2}}$ & 0 & 0 & $-{2 \over 3\sqrt{2}}$  \\ 
$B^+ \rightarrow K^{*0} \rho^+$ & 0 & 0 & 0 & 0 & 0 & 1 & 0 & 0 & 0 & 0 & 0 & $-{1 \over 3}$  \\ 
$B^+ \rightarrow K^{*+} \omega$ & 0 & $1 \over \sqrt{2}$ & $1 \over \sqrt{2}$ & 0 & 0 & $1 \over \sqrt{2}$ & $\sqrt{2}$ & 0 & $1 \over 3\sqrt{2}$ & 0 & 0 & $2 \over 3\sqrt{2}$ \\ 
$B^+ \rightarrow K^{*+} \phi $ & 0 & 0 & 0 & 0 & 1 & 0 & 1 & 0 & $-{1 \over 3}$ & 0 & $-{1 \over 3}$ & 0 \\ \hline 
$B^0 \rightarrow K^{*+} \rho^-$ & 0 & $-1$ & 0 & 0 & 0 & $-1$ & 0 & 0 & 0 & 0 & 0 & $-{2 \over 3}$ \\ 
$B^0 \rightarrow K^{*0} \rho^0$ & 0 & 0 & $-{1 \over \sqrt{2}}$ & 0 & 0 & $1 \over \sqrt{2}$ & 0 & 0 & $-{1 \over \sqrt{2}}$ & 0 & 0 & $-{1 \over 3\sqrt{2}}$ \\ 
$B^0 \rightarrow K^{*0} \omega$ & 0 & 0 & $1 \over \sqrt{2}$ & 0 & 0 & $1 \over \sqrt{2}$ & $\sqrt{2}$ & 0 & $1 \over 3\sqrt{2}$ & 0 & 0 & $-{1 \over 3\sqrt{2}}$  \\ 
$B^0 \rightarrow K^{*0} \phi$ & 0 & 0 & 0 & 0 & 1 & 0 & 1 & 0 & $-{1 \over 3}$ & 0 & $-{1 \over 3}$ & 0 
\end{tabular}
\vspace{2cm} 
\caption{Coefficients of SU(3)$_F$ breaking effects in $B \rightarrow V_1 V_2$ ( $|\Delta S| = 1$ ).}
\begin{tabular}{|c||c|c|c|c|c|c|c|c|} 
 $B \rightarrow V_1 V_2$ & $T^{\prime V_2}_1$ & $C^{\prime V_1}_1$ & $P^{\prime V_1}_{(3)}$ & $P^{\prime V_2}_{1}$ & $S^{\prime V_1}_{1 (3)}$ & $P_{EW,1(3)}^{\prime V_1}$ & $P_{EW,(3)}^{\prime C,V_1}$ & $P_{EW,1}^{\prime C,V_2}$ \\ \hline \hline
$B^+ \rightarrow K^{*+} \rho^0$ & $-{1 \over \sqrt{2}}$ & $-{1 \over \sqrt{2}}$ & 0 & $-{1 \over \sqrt{2}}$ & 0 & $-{1 \over \sqrt{2}}$ & 0 & $-{2 \over 3\sqrt{2}}$  \\ 
$B^+ \rightarrow K^{*0} \rho^+$ & 0 & 0 & 0 & 1 & 0 & 0 & 0 & $-{1 \over 3}$  \\ 
$B^+ \rightarrow K^{*+} \omega$ & $1 \over \sqrt{2}$ & $1 \over \sqrt{2}$ & 0 & $1 \over \sqrt{2}$ & $\sqrt{2}$ & $1 \over 3\sqrt{2}$ & 0 & $2 \over 3\sqrt{2}$ \\ 
$B^+ \rightarrow K^{*+} \phi $ & 0 & 0 & (1) & 0 & (1) & ($-{1 \over 3}$) & ($-{1 \over 3}$) & 0 \\ \hline 
$B^0 \rightarrow K^{*+} \rho^-$ & $-1$ & 0 & 0 & $-1$ & 0 & 0 & 0 & $-{2 \over 3}$ \\ 
$B^0 \rightarrow K^{*0} \rho^0$ & 0 & $-{1 \over \sqrt{2}}$ & 0 & $1 \over \sqrt{2}$ & 0 & $-{1 \over \sqrt{2}}$ & 0 & $-{1 \over 3\sqrt{2}}$ \\ 
$B^0 \rightarrow K^{*0} \omega$ & 0 & $1 \over \sqrt{2}$ & 0 & $1 \over \sqrt{2}$ & $\sqrt{2}$ & $1 \over 3\sqrt{2}$ & 0 & $-{1 \over 3\sqrt{2}}$  \\ 
$B^0 \rightarrow K^{*0} \phi$ & 0 & 0 & (1) & 0 & (1) & ($-{1 \over 3}$) & ($-{1 \over 3}$) & 0 
\end{tabular}
\newpage
\caption{Absolute values (in units of $10^{-4}$ GeV) of SU(3)$_F$ amplitudes averaged over helicity states for different values of $\xi (= 1/ N_c)$ in $B \rightarrow VV$ decays.  The absolute values are estimated in the factorization approximation, using the form factors in the BSW model [Lattice QCD and Light-cone QCD sum rules].}
\begin{tabular}{|c||c|c|c|c|}
\mbox{}   & $\xi =0$ & $\xi =0.2$ & $\xi =1/3$ & $\xi =0.5$  \\ \hline \hline
$|T^{V}|$  & 6.3 [6.7] & 6.0 [6.3] & 5.7 [6.0] & 5.3 [5.7]    \\
$|C^{V}|$  & 1.8 [1.9] & 0.53 [0.57] & 0.31 [0.33]  & 1.4 [1.4]  \\
$|P^{V}|$  & 0.73 [0.77] & 0.67 [0.70] & 0.63 [0.67] & 0.57 [0.60]  \\
$|S^{V}|$  & 0.57 [0.60] & 0.23 [0.24] & 0.0043 [0.0043] & 0.27 [0.29] \\
$|P^{V}_{EW}|$ & 0.16 [0.16] & 0.15 [0.16] & 0.14 [0.15] & 0.14 [0.14] \\
$|P^{C,V}_{EW}|$ & 0.031 [0.032] & 0.0004 [0.0005] & 0.021 [0.022] & 0.047 [0.050]  \\ \hline
$|T^{\prime V}|$ & 1.5 [1.5] & 1.4 [1.4] & 1.3 [1.4] & 1.2 [1.3]   \\
$|C^{\prime V}|$ & 0.40 [0.43] & 0.12 [0.13] & 0.070 [0.073] & 0.31 [0.33]  \\
$|P^{\prime V}|$ & 4.0 [4.3] & 3.7 [4.0] & 3.3 [3.7] & 3.3 [3.3]  \\
$|S^{\prime V}|$ & 3.1 [3.3] & 1.3 [1.3] & 0.021 [0.023] & 1.5 [1.6]  \\
$|P^{\prime V}_{EW}|$ & 0.83 [0.87] & 0.77 [0.83] & 0.77 [0.80] & 0.73 [0.77]  \\
$|P^{\prime C,V}_{EW}|$ & 0.16 [0.17] & 0.0026 [0.0028] & 0.11 [0.12] & 0.25 [0.26]
\end{tabular}
\newpage
\caption{Absolute values (in units of $10^{-4}$ GeV) of SU(3)$_F$ breaking effects averaged over helicity states for different values of $\xi (= 1/ N_c)$ in $B \rightarrow VV$ decays.  The absolute values are estimated in the factorization approximation, using the form factors in the BSW model [Lattice QCD and Light-cone QCD sum rules].}
\begin{tabular}{|c||c|c|c|c|}
\mbox{}   & $\xi =0$ & $\xi =0.2$ & $\xi =1/3$ & $\xi =0.5$  \\ \hline \hline
$|P^{V}_3|$ & 0.15 [0.23] & 0.14 [0.21] & 0.13 [0.20] & 0.12 [0.18]  \\
$|S^{V}_3|$ & 0.097 [0.097] & 0.040 [0.040] & 0.0007 [0.0007] & 0.047 [0.047]  \\
$|P^{V}_{EW,3}|$ & 0.027 [0.027] & 0.026 [0.026] & 0.025 [0.025] & 0.024 [0.024]  \\
$|P^{C,V}_{EW,3}|$ & 0.0063 [0.0093] & 0.0001 [0.0001] & 0.0043 [0.0063] & 0.010 [0.014]  \\ \hline
$|T^{\prime V}_1|$ & 0.067 [0.067] & 0.063 [0.063] & 0.060 [0.060] & 0.057 [0.057]  \\
$|C^{\prime V}_1|$ & 0.067 [0.11] & 0.020 [0.031] & 0.011 [0.018] & 0.050 [0.080] \\
$|P^{\prime V}_{1}|$ & 0.19 [0.19] & 0.18 [0.17] & 0.17 [0.16] & 0.15 [0.15]  \\
$|P^{\prime V}_{(3)}|$ & (1.5) [(2.0)] & (1.4) [(1.8)] & (1.3) [(1.7)] & (1.2) [(1.5)]  \\
$|S^{\prime V}_{1}|$ & 0.50 [0.77] & 0.20 [0.32] & 0.0033 [0.0053] & 0.24 [0.37]  \\
$|S^{\prime V}_{(3)}|$ & (1.1) [(1.5)] & (0.43) [(0.57)] & (0.0073) [(0.010)] & (0.53) [(0.70)]  \\
$|P^{\prime V}_{EW,1}|$ & 0.13 [0.21] & 0.12 [0.20] & 0.12 [0.19] & 0.12 [0.18]  \\
$|P^{\prime V}_{EW,(3)}|$ & (0.29) [(0.37)] & (0.28) [(0.37)] & (0.27) [(0.33)] & (0.26) [(0.33)]  \\
$|P^{\prime C,V}_{EW,1}|$ & 0.0077 [0.0073] & 0.0001 [0.0001] & 0.0050 [0.0050] & 0.011 [0.011]  \\
$|P^{\prime C,V}_{EW,(3)}|$ & (0.057) [(0.073)] & (0.0009) [(0.0012)] & (0.040) [(0.053)] & (0.087) [(0.11)]
\end{tabular}
\end{table}

\newpage
\begin{figure}[htb]
\vspace{1 cm}

\centerline{ \DESepsf(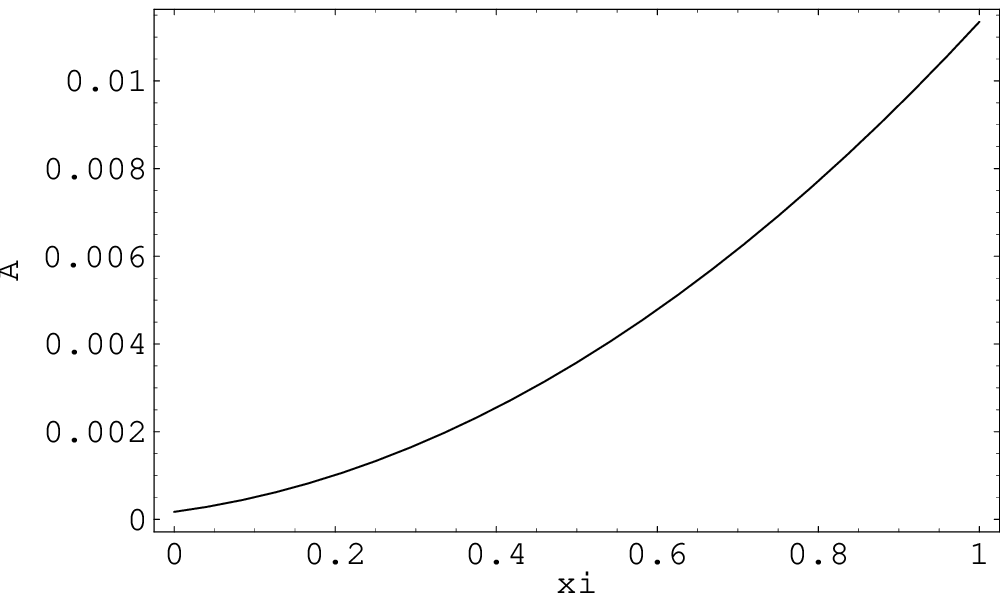 width 12 cm) }
\vspace{1 cm}
\caption {Absolute value squared of sum of the effective coefficients for $b \rightarrow s$, $A \equiv | a_3 +a_4 +a_5 |^2$, as a funtion of $\xi$.  The penguin amplitudes $P^{\prime V}$ and $S^{\prime V}$ are proportional to $a_4$ and $(a_3 +a_5)$, respectively.  Thus, in factorization approximation the branching ratio for $B \rightarrow K^* \phi$ is roughly proportional to $A$ (if one neglects relatively small contribution from electroweak penguins).  It has been shown that in $B \rightarrow K^* \phi$ the experimentally favored values of $\xi$ are $0.4 \leq \xi \leq 0.6$. }
\end{figure}


\newpage
\begin{thebibliography}{[001]}
\bibitem{1} M.S. Alam \emph{et al.} (CLEO Collaboration), CLEO CONF 97-23, 1997; 
A. Anastassov \emph{et al.} (CLEO Collaboration), CLEO CONF 97-24, 1997; R. Godang \emph{et al.}, (CLEO Collaboration), Phys. Rev. Lett. {\bf 80}, 3456 (1998); B.H. Behrens \emph{et al.}, (CLEO Collaboration), \emph{ibid.} {\bf 80}, 3710 (1998); J.G. Smith, Report No. COLO-HEP-395, 1998;    
For a recent review, see K. Lingel, T. Skwarnicki, and J.G. Smith, Annu. Rev. Nucl. Part. Sci. (to be published), hep-ex/9804015. 
\bibitem{2} T. Bergfeld \emph{et al.}, (CLEO Collaboration), Phys. Rev. Lett. {\bf 81} (1998) 272.       
\bibitem{3} L.-L. Chau, H.-Y. Cheng, W.K. Sze, H. Yao, and B. Tseng, Phys. Rev. D {\bf 43} (1991) 2176, and the earlier references quoted therein. 
\bibitem{3a} D. Du and Z. Xing, Phys. Lett. B {\bf 312} (1993) 199;
H.-Y. Cheng and B. Tseng, Phys. Lett. B {\bf 415} (1997) 263; Report No. IP-ASTP-04-97/NTU-TH-97-09, hep-ph/9708211; Phys. Rev. D {\bf 58} (1998) 094005. 
\bibitem{300} A. DeAndrea, N. Di Bartolomeo, R. Gatto, and G. Nardulli, Phys. Lett. B {\bf 318} (1993) 549; A. DeAndrea, N. Di Bartolomeo, R. Gatto, F. Feruglio, and G. Nardulli, \emph{ibid.} {\bf 320} (1994) 170. 
\bibitem{301} G. Kramer and W.F. Palmer, Phys. Rev. D {\bf 52} (1995) 6411; Z. Phys. C {\bf 66} (1995) 429. 
\bibitem{302} M. Ciuchini, E. Franco, G. Martinelli, and L. Silvestrini, Nucl. Phys. B {\bf 501} (1997) 271; Erratum - \emph{ibid.} B {\bf 531} (1998) 656; M. Ciuchini, R. Contino,  E. Franco, G. Martinelli, and L. Silvestrini, Nucl. Phys. B {\bf 512} (1998) 3. 
\bibitem{303} N.G. Deshpande and X.-G. He, Phys. Lett. B {\bf 336} (1994) 471; Phys. Rev. Lett. {\bf 74} (1995) 26;  R. Fleischer, Z. Phys. C {\bf 62} (1994) 81; A.J. Buras and R. Fleischer, Phys. Lett. B {\bf 365} (1996) 390.      
\bibitem{4} N.G. Deshpande, B. Dutta, and Sechul Oh, Phys. Lett. B (to be published), hep-ph/9712445. 
\bibitem{5} G. Valencia, Phys. Rev. D {\bf 39} (1989) 3339; A.N. Kamal and C.W. Luo, Phys. Lett. B {\bf 388} (1996) 633.  
\bibitem{500} G. Kramer and W.F. Palmer, Phys. Rev. D {\bf 45} (1992) 193; Phys. Lett. B {\bf 279} (1992) 181; Nucl. Phys. B {\bf 428} (1994) 77; 
\bibitem{6} A. Ali and C. Greub, Phys. Rev. D {\bf 57} (1998) 2996; 
N.G. Deshpande, B. Dutta, and Sechul Oh, Phys. Rev. D {\bf 57} (1998) 5723. 
\bibitem{7} A. Ali, G. Kramer, and Cai-Dian L\"{u}, Phys. Rev. D {\bf 58} (1998) 094009, and references quoted therein.  
\bibitem{8} M. Gronau, O.F. Hern\'{a}ndez, D. London, and J.L. Rosner, Phys. Rev. D {\bf 50} (1994) 4529. 
\bibitem{9} M. Gronau, O.F. Hern\'{a}ndez, D. London, and J.L. Rosner, Phys. Rev. D {\bf 52} (1995) 6356. 
\bibitem{9a} M. Gronau, O.F. Hern\'{a}ndez, D. London, and J.L. Rosner, Phys. Rev. D {\bf 52} (1995) 6374; A.S. Dighe, M. Gronau, and J.L. Rosner, Phys. Lett. B {\bf 367} (1996) 357; Erratum - \emph{ibid.} {\bf 377} (1996) 325; A.S. Dighe, Phys. Rev. D {\bf 54} (1996) 2067.  
\bibitem{10} A.S. Dighe, M. Gronau, and J.L. Rosner, Phys. Rev. D {\bf 57} (1998) 1783.
\bibitem{11} M. Neubert and B. Stech, hep-ph/9705292, to appear in \emph{Heavy Flavors}, 2nd ed., edited by A.J. Buras and M. Lindner (World Scientific, Singapore, 1998). 
\bibitem{12} A.S. Dighe, I. Dunietz, H.J. Lipkin, and J.L. Rosner, Phys. Lett. B {\bf 369} (1996) 144; A.S. Dighe and \'{S}. Sen, Report No. IC/98/177, hep-ph/9810381. 
\bibitem{1200} D. Atwood and A. Soni, Phys. Lett. B {\bf 405} (1997) 150; W.-S. Hou and B. Tseng, Phys. Rev. Lett. {\bf 80} (1998) 434; A. Datta, X.-G. He, and S. Pakvasa, Phys. Lett. B {\bf 419} (1998) 369; A.L. Kagan and A.A. Petrov, Report No. UCHEP-27/UMHEP-443, hep-ph/9707354; H. Fritzsch, Phys. Lett. B {\bf 415} (1997) 83. 
\bibitem{annih} It has been pointed out in Ref.\cite{7} that the annihilation contributions in $B^+ \rightarrow K^{*+} \phi$ and $K^{*+} \bar K^{*0}$ may be large.  However in factorization the annihilation contributions introduce more untested form factors which are model-dependent.  
\bibitem{13} M. Bauer and B. Stech, Phys. Lett. B {\bf 152} (1985) 380; M. Bauer, B. Stech, and M. Wirbel, Z. Phys. C {\bf 34} (1987) 103.  
\bibitem{14} A. Ali, V.M. Braun, and H. Simma, Z. Phys. C {\bf 63} (1994) 437; 
J.M. Flynn \emph{et el.} (UKQCD Collaboration), Nucl. Phys. B {\bf 461} (1996) 327; 
L. Del Debbio \emph{et el.} (UKQCD Collaboration), Phys. Lett. B {\bf 416} (1998) 392. 
  
\end{thebibliography}
\end{document}